\documentclass{article}
\usepackage{spconf,amsmath,graphicx,hyperref}
\usepackage{xcolor}
\usepackage{amssymb}
\usepackage{booktabs} % 用于美化表格
\usepackage{array}    % 用于调整列格式
\usepackage{multirow}
\usepackage[table]{xcolor}
\usepackage{bm}
\usepackage{cite}
\usepackage{hyperref}

\usepackage{tabularx} % For equal-width columns
\newcolumntype{Y}{>{\centering\arraybackslash}X} % Define a centered X column type

\title{V2A-DPO: Omni-Preference Optimization for Video-to-Audio Generation}
\name{Nolan Chan$^{*,1}$, Timmy Gang$^{*,2}$, Yongqian Wang$^3$, Yuzhe Liang$^4$, Dingdong Wang$^1$\thanks{ $^{*}$ Corresponding author: yjchen@se.cuhk.edu.hk}}
\address{
$^1$ The Chinese University of Hong Kong, Hong Kong SAR, China \\
$^2$ National Research Council Canada, Canada \\
$^3$ The University of Warwick, UK \\
$^4$ Shanghai Jiao Tong University, China
}
\begin{document}
\ninept
\maketitle
\begin{abstract}
This paper introduces V2A-DPO, a novel Direct Preference Optimization (DPO) framework tailored for flow-based video-to-audio generation (V2A) models, incorporating key adaptations to effectively align generated audio with human preferences. Our approach incorporates three core innovations: (1) AudioScore—a comprehensive human preference-aligned scoring system for assessing semantic consistency, temporal alignment, and perceptual quality of synthesized audio; (2) an automated AudioScore-driven pipeline for generating large-scale preference pair data for DPO optimization; (3) a curriculum learning-empowered DPO optimization strategy specifically tailored for flow-based generative models. Experiments on benchmark VGGSound dataset demonstrate that human-preference aligned Frieren and MMAudio using V2A-DPO outperform their counterparts optimized using Denoising Diffusion Policy Optimization (DDPO) as well as pre-trained baselines. Furthermore, our DPO-optimized MMAudio achieves state-of-the-art performance across multiple metrics, surpassing published V2A models. Our demos are available at \href{https://seujames23.github.io/V2A-DPO/}{https://seujames23.github.io/V2A-DPO/}.
\end{abstract}
\begin{keywords}
Video-to-Audio, Direct Preference Optimization, Human Preference Alignment, Flow Matching, Curriculum Learning 
\end{keywords}
\vspace{-0.4cm} 
\section{Introduction}
\label{sec:intro}

Video-to-Audio Generation (V2A) aims to synthesize well-aligned audio conditioned on video features with optional text prompt, which usually takes the semantic and temporal information into account. By matching silent videos with high-quality and semantically consistent audios, V2A models complement modern video generation systems, which predominantly concentrate on visual synthesis \cite{xu2024video}.

Recent years have witnessed significant progress in V2A models. Early GAN-based models \cite{chen2020generating,su2020audeo,ghose2022foleygan} utilize the generators for audio generation from visual features and the discriminators to distinguish the generated audio from ground-truth. Recent breakthroughs in transformer-based autoregressive models offer powerful capabilities in V2A tasks \cite{viertola2025temporally,gan2020foley,liu2025thinksound,sheffer2023hear,mei2024foleygen,liang2025deepsound}. Among them, ThinkSound \cite{liu2025thinksound} and DeepSound-V1 \cite{liang2025deepsound} harness Chain-of-Thought (CoT) reasoning to enable stepwise, interactive generation and editing for V2A. Additionally, other cutting-edge approaches \cite{xu2024video,chen2025video,mo2025foley,luo2023diff,chen2024action2sound,wang2024frieren,wang2024v2a,zhang2024foleycrafter,cheng2025mmaudio} have introduced novel diffusion- or flow matching-based models to address this task and achieve notable performance.

Despite of the rapid development, prior V2A models still exhibit several notable limitations. First, their style control is restricted to video-audio pairs used for training, limiting the flexibility and precision of stylistic variations \cite{chen2020generating,su2020audeo,ghose2022foleygan,viertola2025temporally,gan2020foley,liu2025thinksound,sheffer2023hear,mei2024foleygen,liang2025deepsound,xu2024video,chen2025video,mo2025foley,luo2023diff,chen2024action2sound,wang2024frieren,wang2024v2a,zhang2024foleycrafter,cheng2025mmaudio}. When given a scenario with significant differences from the training data during test-time, existing models often generate audios with inappropriate styles. Second, the aesthetic quality of synthesized audio remains challenging to assess through explicit reward modeling—a critical aspect that most V2A approaches overlook \cite{chen2020generating,su2020audeo,ghose2022foleygan,viertola2025temporally,gan2020foley,liu2025thinksound,sheffer2023hear,mei2024foleygen,liang2025deepsound,xu2024video,chen2025video,mo2025foley,luo2023diff,chen2024action2sound,wang2024frieren,wang2024v2a,zhang2024foleycrafter,cheng2025mmaudio}. However, even when a generated audio is semantically relevant and temporally aligned, it may still fail to provide a sense of immersion for the listener due to its lack of aesthetic quality. Third, previous approaches typically employ isolated quantitative metrics to assess the semantic and temporal alignment or perceptual quality of generated audio in a separate manner \cite{chen2020generating,su2020audeo,ghose2022foleygan,viertola2025temporally,gan2020foley,liu2025thinksound,sheffer2023hear,mei2024foleygen,liang2025deepsound,xu2024video,chen2025video,mo2025foley,luo2023diff,chen2024action2sound,wang2024frieren,wang2024v2a,zhang2024foleycrafter,cheng2025mmaudio}. However, the absence of a comprehensive scoring system that holistically integrates multiple metrics imposes significant limitations on accurately assessing the performance of V2A models.

To address these shortcomings, this paper introduces V2A-DPO, a novel Direct Preference Optimization (DPO) \cite{rafailov2023direct} framework tailored for flow-based V2A generation models, incorporating key adaptations to effectively align generated audio with human preferences. Specifically, we first propose AudioScore, a comprehensive human preference-aligned scoring system designed to simultaneously assess semantic consistency, temporal alignment, and perceptual quality of generated audio. By integrating a small set of human-annotated preference pairs with automatically generated preference pairs through an AudioScore-driven pipeline, we then efficiently build a substantial dataset suitable for DPO optimization. By incorporating a curriculum learning \cite{bengio2009curriculum} paradigm into the DPO optimization process, our human preference-aligned V2A models significantly enhance the perceptual quality, semantic consistency, temporal alignment, and even aesthetic appeal of the generated audio. Experiments on benchmark VGGSound dataset suggest that human-preference aligned Frieren \cite{wang2024frieren} and MMAudio \cite{cheng2025mmaudio} using V2A-DPO outperform their counterparts optimized using Denoising Diffusion Policy Optimization (DDPO) \cite{black2023training} as well as pre-trained baselines, achieving significant improvements of up to \textbf{1.81 absolute} (\textbf{10.4\% relative}) in IS and \textbf{0.86 absolute} (\textbf{2.6\% relative}) in IB-score, along with a reduction in DeSync of \textbf{0.09 absolute} (\textbf{20.5\% relative}). Furthermore, our DPO-optimized MMAudio achieves state-of-the-art performance across multiple metrics, surpassing published V2A models.

Our main contributions are summarized as follow:

(a) We pioneer the adaptation of DPO to flow-based V2A models, addressing the unique challenges of aligning audio generation outputs with human preferences.

(b) We introduce several key adaptations to the DPO optimization framework, including: (1) AudioScore—a comprehensive human preference-aligned scoring system for assessing semantic consistency, temporal alignment, and perceptual quality of generated audio; (2) an automated AudioScore-driven pipeline for generating large-scale preference pair data for DPO optimization; (3) a curriculum learning-empowered DPO optimization strategy tailored for flow-based generative models.

(c) To the best of our knowledge, we build the first high-quality video–text prompt–audio preference pair dataset designed for V2A models' alignment with human preferences, simultaneously taking semantic consistency, temporal alignment, perceptual quality, and aesthetic appeal into account.

(d) We validate the proposed V2A-DPO framework through extensive experiments conducted on two open-source pre-trained V2A models. The results demonstrate the robustness and effectiveness of our approach across multiple metrics.

%------------figure1-------------
\begin{figure*}[htbp]
\centering
\setlength{\abovecaptionskip}{0pt plus 1pt minus 3pt}
\includegraphics[scale=0.14]{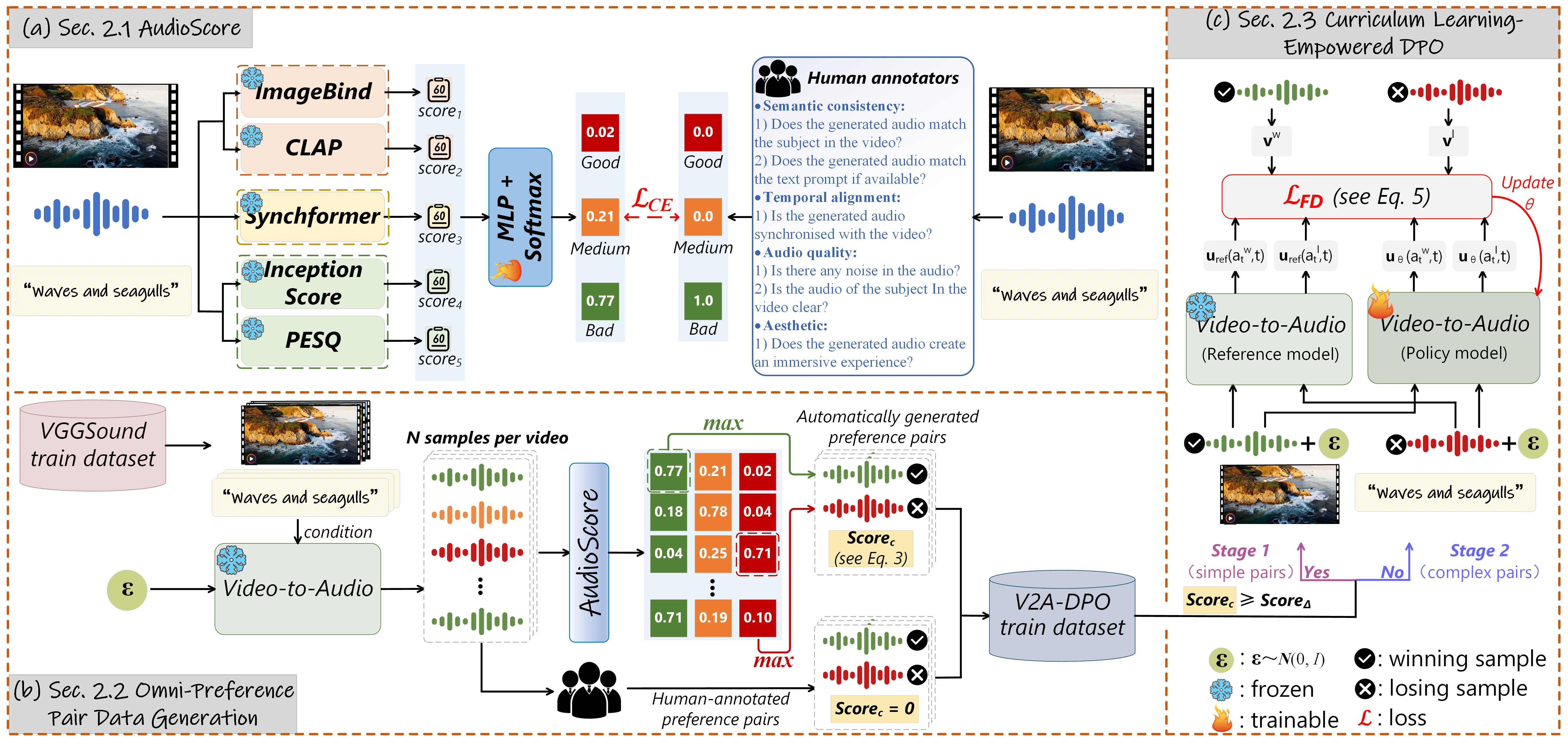}
\vspace{-0.3cm} 
\caption{\textbf{Illustration of the proposed V2A-DPO framework}, including (a) our proposed AudioScore to rate the generated audio with multi-dimensional scores; (b) omni-preference pair data generation combining the automatically generated preference pairs based on AudioScore with a small amount of human-annotated preference pairs; (c) curriculum learning-empowered DPO to optimize V2A models on the complex and simple pairs gradually, which are split according to complexity score $score_c$.}
\label{fig1}
\vspace{-0.3cm}
\end{figure*}
%------------figure1-------------

\vspace{-0.4cm} 
\section{Method}
\label{method}
\vspace{-0.2cm} 
Our proposed V2A-DPO (Omni-Preference Optimization for Video-to-Audio Generation) framework is illustrated in Fig. \ref{fig1}, including a comprehensive human preference-aligned scoring system, an automated preference pair data generation pipeline and a curriculum learning-empowered DPO optimization strategy for V2A models.
\vspace{-0.4cm} 
\subsection{AudioScore}
\label{audioscore}
\vspace{-0.2cm} 
 The quality of generated audio is influenced by multiple factors, including: semantic consistency, temporal alignment, perceptual quality and aesthetic appeal.  Semantic consistency denotes the content consistency between the generated audio and the the video or text prompt. Temporal alignment, on the other hand, focuses on whether the generated video accurately follows the video. Third, perceptual quality includes the clarity and richness of generated audio. Finally, aesthetic appeal mainly focuses on whether generated audio creates an immersive experience for listeners, even when achieving high scores in the first three criteria. To start, we randomly sample 2K videos with text prompts from the VGGSound \cite{vggsound} training set and generate 10K audio samples through pre-trained V2A models that we aim to align. Subsequently, human annotators rate these generated samples and classify them into three categories, i.e. ``Good", ``Medium" and ``Bad", as illustrated in Fig. \ref{fig1}(a).

To address the challenge of the high cost of human annotating, we propose AudioScore, a comprehensive human preference-aligned scoring system, consisting of several frozen-weighted foundation models, MLP and Softmax modules. As shown in Fig. \ref{fig1}(a), AudioScore computes the average cosine similarity between the input visual and generated audio features extracted by ImageBind \cite{girdhar2023imagebind} as video-audio semantic consistency score (IB-score \cite{viertola2025temporally}; $score_1$). Similarly, AudioScore employs CLAP \cite{elizalde2023clap} to measure the semantic consistency score ($score_2$) between the generated audio and the text prompt if available. Additionally, AudioScore uses the synchronization score (DeSync; $score_3$) proposed in \cite{cheng2025mmaudio}, which is predicted by Synchformer \cite{iashin2024synchformer} as the misalignment (in seconds) between audio and video. Furthermore, AudioScore assesses the generation quality using the PANNs-based Inception Score \cite{salimans2016improved} ($score_4$) following \cite{wang2024frieren}, along with the objective metric perceptual evaluation of speech quality score (PESQ; $score_5$) to evaluate the audio quality with the category of human speech. After obtaining five-dimensional score vectors, AudioScore employs two Linear layers with a ReLU between them and a Softmax module to align the automated classification results of the generated audio samples with the human-annotated results using cross entropy loss $\mathcal{L}_{CE}$.
\vspace{-0.2cm} 
\subsection{Omni-Preference Pair Data Generation}
\label{datageneration}
\vspace{-0.2cm} 
To construct the large-scale preference pair dataset, AudioScore is employed in combination with a ``best vs. worst'' selection strategy based on classification probabilities. As shown in Fig. \ref{fig1}(b), for each prompt, our system generates multiple samples using pre-trained V2A models that we aim to align and a preference pair is selected with the highest probability of ``Good'' as the winning sample while the highest probability of ``Bad'' as the losing one. We randomly sample 50K videos with text prompts from the VGGSound training set and then obtain about 46K omni-preference pairs after re-filtering. Given that aesthetic appeal is challenging to assess quantitatively via AudioScore, we combine the automatically generated preference pairs with 2K human-annotated preference pairs, forming a DPO training dataset comprising approximately 48K pairs in total.

\noindent \textbf{Audio generation and scoring.} Given a video $v$ with an optional text prompt $c$, we generate a set of $N$ audios ${a_1,a_2,...,a_N}$, using the pre-trained V2A model. For each generated audio $a_i$, we apply AudioScore $\mathrm{S(\cdot)}$ to score its quality conditioned on the video $v$ with optional text prompt $c$. This scoring system predicts the probability vector $\mathbf{p}_i$ of preference for each audio:
\vspace{-0.2cm}
\begin{small}
\begin{equation}
\begin{aligned}
 \mathbf{p}_i &= \Big(p_i(\mathrm{Good}|a^i,v,c), p_i(\mathrm{Medium}|a^i,v,c), p_i(\mathrm{Bad}|a^i,v,c)\Big) \\
& = \mathrm{S}(a^i,v,c),\  \mathrm{for}\ i=1,2,...,N.
\end{aligned}
\end{equation}
\end{small}

\noindent where, $p_i(*|a^i,v,c)$ refers to the predicted probability of the corresponding category for the $i$th generated audio sample conditioning on video $v$ with optional text prompt $c$ , where $*$ in \{``Good'', ``Medium'', ``Bad''\}.

\noindent \textbf{Preference pair selection.} Among the $N$ generated audios with the corresponding probability vectors {$\mathbf{p}_1,\mathbf{p}_2,...,\mathbf{p}_N$}, we select the audio with the highest probability of “Good” as the winning sample $a^w$ and the video with the highest probability of “Bad” as the losing sample $a^l$. This selection process is formalized as follows:
\vspace{-0.2cm}
\begin{small}
\begin{equation}
\begin{aligned}
(a^w,&a^l)=(a^i,a^j),\ i=arg\ max\ p_i(\mathrm{Good}|a^i,v,c),\\
&j=arg\ max\ p_j(\mathrm{Bad}|a^j,v,c),\ \mathrm{for}\ i,j=1,2,...,N.
\end{aligned}
\end{equation}
\end{small}
\vspace{-0.4cm} 
\subsection{Curriculum Learning-Empowered DPO}
\label{v2adpo}
\vspace{-0.2cm} 
Despite the promising results of DPO in the text-to-speech, text-to-image and text-to-video tasks \cite{gao2025emo,zhang2024speechalign,tian2025preference,majumder2024tango,chen2025diffrhythm+,liu2025videodpo,wallace2024diffusion,huang2025patchdpo,wang2025designdiffusion,na2025boost,Croitoru_2025_CVPR}, randomly sorting the available preference pairs without considering their difficulty during training has been proven suboptimal as minor differences between difficult pairs making it challenging for the models to effectively distinguish \cite{Croitoru_2025_CVPR,liu2025videodpo}. To address this, we introduce a curriculum learning-empowered DPO optimization strategy for V2A models, dividing the training process into two distinct stages based on the complexity scores of the preference pairs. In the first stage, the model is aligned using preference pairs with clearly distinguishable differences, while the second stage utilizes pairs with subtler, more nuanced distinctions. Through curriculum learning, the model is guided to first learn more meaningful alignment cues from simpler preference pairs before progressing to more complex pairs in the subsequent stage. This structured approach enables more stable and gradual improvement of the model’s generative capabilities. 

\noindent \textbf{Probability-based complexity score.} Considering a preference pair $(a^w,a^l)$ with the input video $v$ and optional text prompt $c$, the complexity score of this pair $score_{c}$ can be calculated as Eq. \ref{eq3}. As is shown in Fig. \ref{fig1}(c), when $score_{c}$ exceeds the preset threshold $score_{\Delta}$, the preference pair data will be input into the first stage training; otherwise, the second stage. Note that the complexity scores of 2K human-annotated pairs are all set to zero as we hope that the optimized model will focus on the aesthetic appeal of the generated audio in the second stage.
\vspace{-0.2cm}
\begin{small}
\begin{equation}
\begin{aligned}
score_{c}=\frac{1}2{}\Big[\Big(p^w(\mathrm{Good}|a^w,v,c)-p^l(\mathrm{Good}|a^l,v,c)\Big)\\
+\Big(p^l(\mathrm{Bad}|a^l,v,c)-p^w(\mathrm{Bad}|a^w,v,c)\Big)\Big]
\label{eq3}
\end{aligned}
\end{equation}
\end{small}

\noindent where, $p^w(\mathrm{Good}|a^w,v,c)$ and $p^l(\mathrm{Good}|a^l,v,c)$ denote the probabilities of ``Good" for the winning and losing samples, respectively, while $p^w(\mathrm{Bad}|a^w,v,c)$ and $p^l(\mathrm{Bad}|a^l,v,c)$ represent the probabilities of ``Bad".

\noindent \textbf{Flow-DPO.} Given a fixed dataset $\mathcal{D} = \{(v,c, a_0^w,a_0^l)\}$ where each example contains a video $v$, an optional text prompt $c$ and a pair of audios generated by a reference model $q_\text{ref}$, with $a_0^w$ preferred over $a_0^l$, the reward objective of DPO to optimize the policy model $q_{\theta}$ can be illustrated as following:
\vspace{-0.4cm}
\begin{small}
\begin{equation}
\begin{aligned}
\mathcal{L}_{DPO}(\theta)=-\mathbb{E}_{(v,c,a_0^w,a_0^l)\sim\mathcal{D}}\Big[log\sigma\Big(\beta\,log\frac{q_\theta(a_0^w|v,c)}{q_\text{ref}(a_0^w|v,c)} \\
-\beta\,log\frac{q_\theta(a_0^l|v,c)}{q_\text{ref}(a_0^l|v,c)}\Big)\Big]
\end{aligned}
\end{equation}
\end{small}

\noindent where $\sigma$ is the sigmoid function. $a_0^*$ refers to the target audio sampled from $\mathcal{D}$ in the rectified flow matching with the superscript $*$ indicating either ``$w$'' (for the winning sample) or ``$l$'' (for the losing sample).

In adapting DPO to flow-based models, \cite{liu2025improving} interprets alignment as a classification problem, and optimizes a policy to satisfy human preferences by supervised training. For simplicity, we omit the conditioning video $v$ and optional text prompt $c$ in the following equations. The
Flow-DPO objective $\mathcal{L}_{FD}(\theta)$ is given by:
\vspace{-0.2cm}
\begin{small}
\begin{equation}
\begin{aligned}
\label{eq:diffusion_dpo}
- \mathbb{E} \Big[log\sigma\Big(-\frac{\beta_t}{2}\Big(\|\mathbf{v}^w-\mathbf{u}_\theta(a^w_t,t)\|^2-\|\mathbf{v}^w-\mathbf{u}_\text{ref}(a^w_t,t)\|^2\Big) \\
-\Big(\|\mathbf{v}^l-\mathbf{u}_\theta(a^l_t,t)\|^2-\|\mathbf{v}^l-\mathbf{u}_\text{ref}(a^l_t,t)\|^2\Big)\Big)\Big]
\end{aligned}
\end{equation}
\end{small}

\noindent where $a^*_t=(1-t)a^*_0+t\epsilon,\,\epsilon\sim\mathcal{N}(0,\mathbf{\it{I}})$. And $t\sim\mathcal{U}(0,1)$ is the timestep in rectified flow matching. $\mathbf{u}_\text{ref}$ and $\mathbf{u}_{\theta}$ refer to the predicted vector fields from the policy model $q_{\theta}$ and the reference model $q_\text{ref}$, while $\mathbf{v}^w$ and $\mathbf{v}^l$ denote the target vector fields of the winning and losing samples, respectively. The parameter $\beta_t$ governs the strength of the KL divergence constraint and varies with $t$. In our experiments, we use a constant value $\beta$ in place of $\beta_t$, following the approach in \cite{liu2025improving}, which has been shown to yield btter performance.

Intuitively, minimizing $\mathcal{L}_{FD}(\theta)$ guides the predicted vector field $\mathbf{u}_{\theta}$ closer to the target vector $\mathbf{v}^w$ of the ``preferred'' sample, while pushing it away from $\mathbf{v}^l$ (the ``less preferred'' sample). The strength of this preference signal depends on the differences between the predicted errors and the corresponding reference errors, $\|\mathbf{v}^w - \mathbf{u}_\text{ref}(a_{t}^w, t)\|^2$ and $\|\mathbf{v}^l - \mathbf{u}_\text{ref}(a_{t}^l, t)\|^2$.

\begin{table*}[ht]
\centering
\caption{Video-to-audio results on the VGGSound test set. $\diamond$: the evaluation results in \cite{cheng2025mmaudio} as using the same test set. $\dagger$: reproduced using official evaluation code. $\star$: does not use text prompt during testing.}
\label{table1}
\fontsize{9}{12}\selectfont % sets the font size and line spacing
\setlength{\tabcolsep}{4.3pt} % reduce column spacing
\begin{tabularx}{\textwidth}{ccccccccc}
\toprule
\multirow{2}{*}{\textbf{Sys.}} & \multirow{2}{*}{\textbf{Method}} & \multirow{2}{*}{\textbf{Params}} & \multicolumn{3}{c}{\textbf{Distribution matching}} & \textbf{Audio quality} & \textbf{Semantic align.} & \textbf{Temporal align.} \\
\cmidrule(lr){4-6}
\cmidrule(lr){7-7}
\cmidrule(lr){8-8}
\cmidrule(lr){9-9}
& & & $\bm{\mathrm{FD_{PaSST}}}\downarrow$ & $\bm{\mathrm{KL_{PANNs}}}\downarrow$ & $\bm{\mathrm{KL_{PaSST}}}\downarrow$ & \textbf{IS}$\uparrow$ & \textbf{IB-score}$\uparrow$ & \textbf{DeSync}$\downarrow$ \\
\midrule
1& Seeing\&Hearing $\diamond$ & 415M & 219.01 & 2.26 & 2.30 & 8.58  & 33.99 & 1.20 \\
2& V-AURA $\diamond$ $\star$ & 695M & 218.50 & 2.42 & 2.07 & 10.08 & 27.64 & 0.65 \\
3& FoleyCrafter $\diamond$ & 1.22B & 140.09 & 2.30 & 2.23 & 15.68 & 25.68 & 1.23 \\
4& V2A-Mapper $\diamond$ $\star$ & 229M & 84.57 & 2.69 & 2.56 & 12.47 & 22.58 & 1.23 \\
5& ThinkSound $\dagger$ & 1.30B & 54.92 & \textbf{1.32} & 1.52 & 16.03 & \textbf{34.13} & 0.46 \\
\rowcolor{gray!10}
6& Frieren $\star$ & 159M & 106.10 & 2.73 & 2.86 & 12.25 & 22.78 & 0.85 \\
\rowcolor{gray!10}
7& Frieren-DDPO $\star$ & 159M & 75.41 & 2.58 & 2.61 & 13.12 & 23.19 & 0.65 \\
\rowcolor{gray!10}
8& Frieren-DPO $\star$ & 159M & 69.98 & 2.55 & 2.63 & 13.98 & 24.11 & 0.62 \\
\rowcolor{gray!20}
9& MMAudio & 1.03B & 60.60 & 1.65 & 1.40 & 17.40 & 33.22 & 0.44 \\
\rowcolor{gray!20}
10& MMAudio-DDPO & 1.03B & 54.81 & 1.42 & 1.36 & 18.34 & 33.89 & 0.39 \\
\rowcolor{gray!20}
11& MMAudio-DPO & 1.03B & \textbf{51.38} & 1.38 & \textbf{1.34} & \textbf{19.21} & 34.08 & \textbf{0.35} \\
\bottomrule
\end{tabularx}
\end{table*}
%----table1------

\vspace{-0.4cm}
\section{Experiments}
\label{experiments}
\vspace{-0.2cm} 
\subsection{Experimental setup}
\label{setup}
\vspace{-0.2cm} 
\noindent \textbf{Datasets.} As detailed in Sec. \ref{datageneration}, all our experiments are performed on the constructed omni-preference pair data based on the VGGSound \cite{vggsound} dataset, which contains a class label (310 classes in total) for each video. For a fair comparison, we evaluate our models on the same test set of MMAudio \cite{cheng2025mmaudio} due to data contamination. 

\noindent \textbf{Generative models.} We conduct our experiments by using two pre-trained V2A flow-based models: MMAudio-L-44.1kHz \cite{cheng2025mmaudio} and Frieren \cite{wang2024frieren} with the parameters of 1.03B and 159M, respectively.

\noindent \textbf{Baselines.} We compare our optimized V2A models against five state-of-the-art models: diffusion-based Seeing\&Hearing \cite{xing2024seeing}, FoleyCrafter \cite{zhang2024foleycrafter},  V2A-Mapper \cite{wang2024v2a}, and autoregression-based V-AURA \cite{viertola2025temporally}, Thinksound \cite{liu2025thinksound}. Additionally, we compare DPO  with another reinforcement learning method, namely DDPO \cite{black2023training}, with the probability of ``Good'' predicted by AudioScore as the reward.

%----table2------
\begin{table}[ht]
\vspace{-0.5cm}
\centering
\caption{Ablation study on the effect of the KL divergence constraint parameter $\beta$ and preset threshold $score_{\Delta}$ on the performance of optimized MMAudio using V2A-DPO.}
\label{table2}
\fontsize{9}{11}\selectfont % sets the font size and line spacing
\setlength{\tabcolsep}{3pt} % reduce column spacing
\begin{tabularx}{\columnwidth}{ccccccc}
\toprule
\textbf{Sys.} & $\bm{\beta}$ & $\bm{score_{\Delta}}$ & $\bm{\mathrm{FD_{PaSST}}}\downarrow$ & \textbf{IS}$\uparrow$ & \textbf{IB-score}$\uparrow$ & \textbf{DeSync}$\downarrow$ \\
\midrule
\rowcolor{gray!10}
0 & 600 & 0.7(81\%) & \textbf{51.38} & 19.21 & \textbf{34.08} & \textbf{0.35} \\
1 & 1000 & \multirow{4}{*}{0.7(81\%)} & 82.36 & 11.15 & 24.88 & 0.43 \\
2 & 800 & & 64.85 & 17.33 & 29.91 & 0.38 \\
3 & 400 & & 53.46 & \textbf{19.46} & 33.92 & 0.35 \\
4 & 200 & & 55.81 & 18.79 & 33.93 & 0.36 \\
\midrule
5 & \multirow{4}{*}{600} & 1.0(0\%) & 52.66 & 18.83 & 33.79 & 0.36 \\
6 & & 0.9(8\%) & 52.58 & 18.72 & 33.90 & 0.36 \\
7 & & 0.8(45\%) & 51.96 & 19.02 & 33.98 & 0.35 \\
8 & & 0.6(89\%) & 51.63 & 19.10 & 34.01 & 0.35 \\
\bottomrule
\end{tabularx}
\vspace{-0.2cm} 
\end{table}
%----table2------

\noindent \textbf{Evaluation metrics.} We assess the generation quality using the same metrics in \cite{cheng2025mmaudio}. Specifically, we use Fréchet Distance with PaSST \cite{koutini2021efficient} ($\mathrm{FD_{PaSST}}$)
as embedding models and Kullback–Leibler distance with PANNs ($\mathrm{KL_{PANNs}}$) and PaSST ($\mathrm{KL_{PaSST}}$) as classifiers following \cite{liu2023audioldm} to assess the similarity in feature distribution between ground-truth audio and generated audio. Moreover, we use the Inception Score \cite{salimans2016improved}, IB-score \cite{viertola2025temporally}, and DeSync \cite{cheng2025mmaudio} to assess perceptual quality, semantic consistency, and temporal alignment.

\noindent \textbf{Implementation details.} All settings of optimized models are identical to corresponding pre-trained models, i.e. MMAudio-L-44.1kHz and Frieren. We train V2A models for 12K steps with a global batch size of 8, using the AdamW optimizer with a learning rate of 5e-6 and linear warmup steps of 1K. For each prompt, the number of generated audios $N$ is set to 5. Additionally, the KL divergence
constraint parameter $\beta$ and preset threshold $score_{\Delta}$ are set to 600 and 0.7, respectively. During inference, we assign the guidance scale $\gamma$ in the classifier-free guidance (CFG) \cite{ho2022classifier} to 4.5. All experiments are conducted on 8 NVIDIA A100 GPUs.

\vspace{-0.5cm} 
\subsection{Experimental results}
\label{results}
\begin{figure}[htbp]
\vspace{-0.2cm}
\centering
\setlength{\abovecaptionskip}{0pt plus 1pt minus 3pt}
\includegraphics[scale=0.15]{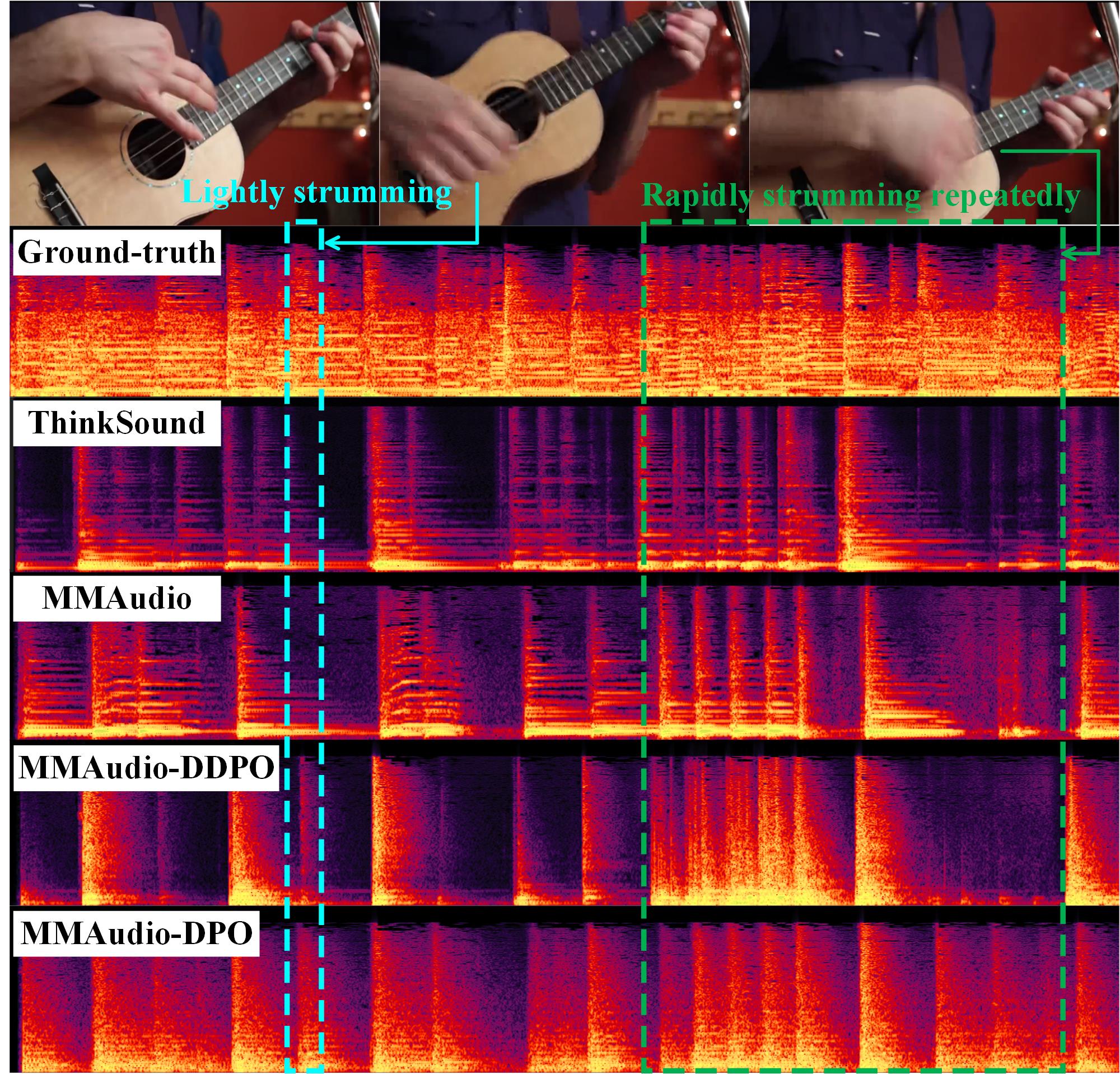}
\caption{Illustration of generation performance of V2A models.}
\label{fig2}
\vspace{-0.4cm}
\end{figure}

\vspace{-0.2cm} 
\noindent \textbf{Performance comparison between DPO-, DDPO-optimized and base models.} As shown in Tab. \ref{table1}, the human-preference aligned MMAudio using V2A-DPO (Sys.11) consistently outperforms the DDPO-optimized (Sys.10) and pre-trained (Sys.9) models in audio quality and semantic alignment, with significant increases in IS and IB-score by up to 1.81 and 0.86 absolute (10.4\% and 2.6\% relative; Sys.11 vs. Sys.9), respectively. Additionally, the temporal alignment performance of the optimized MMAudio has been significantly improved, with a decrease in DeSync reaching 0.09 absolute (20.5\% relative; Sys.11 vs. Sys.9). As intuitively illustrated in Fig. \ref{fig2}, the DPO-optimized MMAudio can align with both the hand movements of ``lightly strumming'' (blue dotted lines) and ``rapidly strumming repeatedly'' (green dotted lines) in the demo, while the DDPO-optimized and pre-trained MMAudio models fail. The same trend can be found in Frieren optimization process (Sys.8 vs. Sys.6,7).
%------------figure2-------------
%------------figure2-------------

\noindent \textbf{Performance comparison between the DPO-optimized and published state-of-the-art V2A models.} The optimized MMAudio using V2A-DPO obtains better performance accross multiple metrics than other models in Tab. \ref{fig1} (Sys.11 vs. Sys.1-5,6,9), except for $\mathrm{KL_{PANNs}}$ and IB-score. We hypothesize that this discrepancy may be attributable to ThinkSound's CoT to analyze visual cues and improve semantic alignment. More demos are available on our \href{https://nolanchan23.github.io/V2A-DPO/}{website}.

\noindent \textbf{Ablation study.} Tab. \ref{table2} shows the impact of the KL divergence constraint parameter $\beta$ and preset threshold $score_{\Delta}$ on the performance of DPO-optimized MMAudio. We observe that the optimized model can generate more semantically consistent and better temporally aligned audios when $\beta$ equals to 600. Furthermore, the performance of DPO-optimized models varies with $score_\Delta$, as different proportions of simple pairs used in the first stages. Note that the curriculum learning-empowered DPO degrades into a regular DPO with a significant decrease in model performance.

\vspace{-0.4cm} 
\section{Conclusion}
\label{conclusion}
\vspace{-0.3cm} 
We introduce V2A-DPO, a novel DPO framework tailored for flow-based V2A models to effectively align generated audio with human preferences. Our approach incorporates three key adaptations: (1) AudioScore; (2) an automated preference pair data generation pipeline; (3) a curriculum learning-empowered DPO optimization strategy. Experiments on the VGGSound dataset demonstrate that optimized Frieren and MMAudio using V2A-DPO outperform DDPO-optimized and pre-trained baselines. Furthermore, our DPO-optimized MMAudio achieves state-of-the-art performance across multiple metrics, surpassing published V2A models.

\bibliographystyle{IEEEbib}
\bibliography{strings,refs}

\end{document}